\documentclass[twocolumn,showpacs,aps,floatfix,superscriptaddress]{revtex4}
\usepackage{amsmath,amssymb,eucal,graphicx}
\usepackage{amsfonts}

\begin{document}
\title{The Simplest Piston Problem II: Inelastic Collisions}
\author{Pablo I. Hurtado}
\email{phurtado@buphy.bu.edu}
\affiliation{Institute \emph{Carlos I} for Theoretical and Computational Physics, \\
Universidad de Granada, 18071 Granada, Spain}
\affiliation{Department of Physics, Boston University, Boston, Massachusetts 02215, USA}
\author{S.~Redner}
\email{redner@bu.edu}
\affiliation{Theoretical Division and Center for Nonlinear Studies, Los Alamos National Laboratory,
Los Alamos, New Mexico 87545, USA}
\altaffiliation{Permanent address: Department of Physics, Boston University, 
Boston, Massachusetts 02215, USA}

\begin{abstract}
  
  We study the dynamics of three particles in a finite interval, in which two
  light particles are separated by a heavy ``piston'', with elastic
  collisions between particles but inelastic collisions between the light
  particles and the interval ends.  A symmetry breaking occurs in which the
  piston migrates near one end of the interval and performs small-amplitude
  periodic oscillations on a logarithmic time scale.  The properties of this
  dissipative limit cycle can be understood simply in terms of a effective
  restitution coefficient picture.  Many dynamical features of the three-particle
  system closely resemble those of the many-body inelastic piston problem.

\end{abstract}
\maketitle

\section{INTRODUCTION}

In the preceding paper, denoted as HR \cite{HR1}, we discussed the collision
dynamics of an elastic three-particle system on a finite interval that consists
of a massive particle---a piston---that separates two lighter particles.  The
motivation for studying this idealized system was to shed light on the
enigmatic piston problem \cite{C60}, where a gas-filled container is divided
into two compartments by a heavy but freely moving piston.  When the gases in
each compartment have different initial thermodynamic states and when the
piston moves without friction, the approach to equilibrium is unexpectedly
complex and still incompletely understood \cite{L99,KVM00,GPL02,CLS02}.

As discussed in HR, some of the rich phenomenology of the piston problem can
be captured by the much simpler three-particle system in a finite interval.  To
understand the evolution of the latter system, it proved convenient to map
the trajectories of the 3 particles on the line onto an equivalent elastic
billiard particle that moves within a highly skewed tetrahedral region, with
the specular reflection whenever the billiard hits the tetrahedron boundaries
\cite{GZ,KT,T,G,R04}.  From this simple geometrical mapping, we deduced
several anomalous dynamical properties of the three-particle system, such as the
power-law distribution of time intervals for the piston to make successive
crossings of the interval midpoint.

In the inelastic piston problem, the collisions between the constituent
particles in the gas are inelastic, so that each gas undergoes inelastic
collapse if either the number of particles is sufficiently large or the
restitution coefficient is sufficiently small.  Recent work by Brito \emph{et al.}\ 
\cite{BRV05} has again discovered surprisingly rich dynamics, very different
in character from the elastic case, in which one of the gases cools more
quickly and gets compressed into a solid by the piston.  An even stranger
feature is that this compression is not monotonic, but rather the piston has
superimposed oscillations whose period grows exponentially with time.  Thus
the cooling of the inelastic piston problem is much richer than that of the
classical inelastic gas problem \cite{general}.

Given the complex behavior exhibited by the many-body piston system, we are
again led to investigate a simpler alternative: a three-particle system in
the unit interval that contains a heavy piston that lies between two light
particles.  Collisions between light particles and the ends of the interval
(henceforth termed walls) are inelastic, to mimic the many-particle piston
problem when the gases are inelastic, while the collisions between the
particles and the piston are elastic.

When the light particles have the same initial energy but nonsymmetric
positions, one light particle loses energy more quickly than the other.  As a
consequence, the piston migrates to the wall that is closer to the cooler
light particle.  Somewhat unexpectedly, a typical system eventually falls
into a periodic state on a logarithmic time scale where the piston undergoes
small-amplitude oscillations near one wall with a constant period in $\ln t$,
while the light particles undergo complementary oscillatory motions.  We term
this phenomenon as the log-periodic state.  Another intriguing aspect of the
three-particle system is that it closely mirrors the time evolution in the
many-particle inelastic piston system \cite{BRV05}.  Thus we are able to
understand features of the many-body problem in terms of simple physical
pictures that arise from studying the three-particle system on the interval.

In the next section, we describe the two basic dynamical features of the
three-particle system, namely, the initial symmetry breaking and the log-periodic
state.  We then give a macroscopic description of the collapse process and
the subsequent oscillatory motion of the piston in Sec.~III.  Finally, in
Sec.~IV, we develop an effective restitution coefficient description for the
particle collisions that accounts for many of our observations.  Various
calculational details are given in an appendix.

\section{BASIC PHENOMENOLOGY}

\subsection{Symmetry breaking}

For the many-body system in which the gases on either side of the piston are
inelastic and have identical macroscopic initial conditions, Brito \emph{et al.}~\cite{BRV05} 
found an instability in which one of the gases cools more
rapidly and the piston ultimately compresses the cooler gas into a solid.
While such an instability seems intuitively plausible, an unexpected feature
is that the piston moves nonmonotonically during this cooling, with regular
oscillations that are periodic on a logarithmic time scale.  In this section,
we show that much of this phenomenology also arises in the idealized
three-particle system on the unit interval.

The particles are located at $x_1$, $x_2$, and $x_3$, with $0\leq x_1\leq
x_2\leq x_3\leq 1$.  The light particles, with masses $m_1=m_3=1$ and
locations $x_1$ and $x_3$, collide elastically with a massive piston with
mass $m_2\gg 1$ at $x_2$ and inelastically with the walls.  Thus a light
particle that hits a wall with speed $v=\sqrt{2E}$ is reflected with speed
$rv$, where $r\in [0,1]$ is the restitution coefficient.  The energy change
in this collision is $\Delta E= -E(1-r^2)<0$.

\begin{figure}[t] 
 \vspace*{0.cm}
\includegraphics*[width=0.45\textwidth]{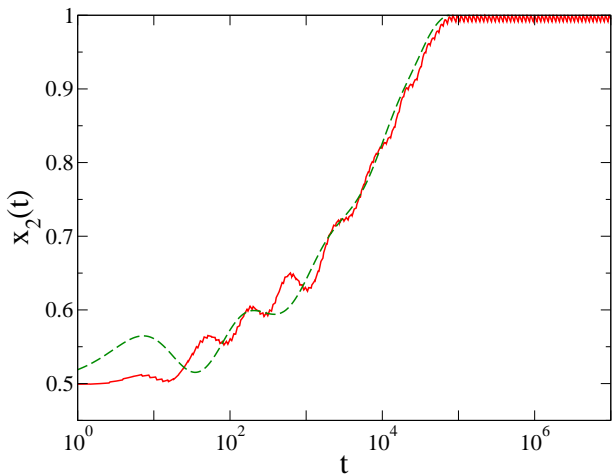}
\caption{(Color online) Piston position $x_2(t)$ versus $t$ on a logarithmic 
  time scale for $m_2=100$ and $r=0.9$.   The solid curve is the simulation
  result while the dashed curve is the prediction from the macroscopic
  equations of motion, Eqs.~(\ref{macro1})--(\ref{macro3}). 
  \label{logosc}}
\end{figure}

\begin{figure}[t] 
 \vspace*{0.cm}
\includegraphics*[width=0.45\textwidth]{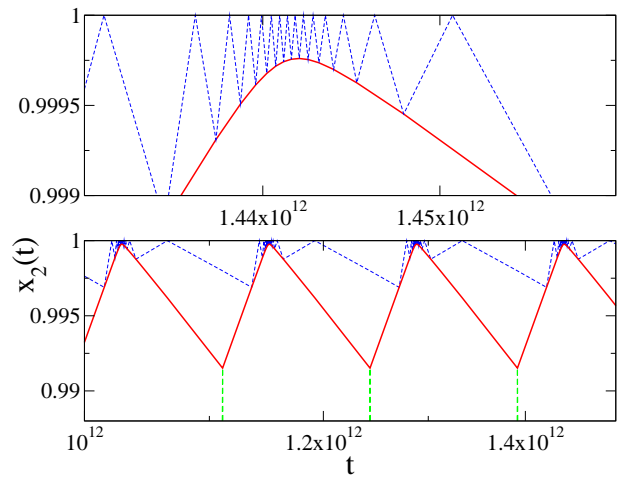}
\caption{(Color online) Magnification of the long time evolution in 
Fig.~\ref{logosc}. Bottom: The log-periodic state. Top: Detail of a 
``rattling'' collision sequence between the piston and 
the trapped light particle.
\label{logfinal}}
\end{figure}

Figure \ref{logosc} shows a representative result for the piston position
$x_2(t)$ versus $t$ on a logarithmic scale for the case $m_2=100$ and
$r=0.9$.  The initial velocities are $(v_1(0),v_2(0),v_3(0))=(1,0,-1)$ so
that two light particles approach the piston with equal and opposite
velocities.  Thus the system initially has zero momentum and total energy
$E=1$.  The initial positions of the light particles were chosen uniformly in
(0,1/2) and in (1/2,1); for the example of Fig.~\ref{logosc},
$(x_1(0),x_2(0),x_3(0))= (0.083\, 25,0.5,0.862\, 83)$.  As a result, the first
collision is between the piston and particle 3.  This small initial asymmetry
eventually drives the piston from oscillations about $x_2=1/2$ to the
nonsymmetric long-time behavior depicted in Fig.~\ref{logosc}.  It bears
emphasizing that the phenomenology of the three-particle system up to
approximately $10^5$ time steps is qualitatively similar to that of the
many-particle inelastic piston problem \cite{BRV05}.

In the long time limit, the piston migrates close to one of the walls.  Which
of the two walls is selected is determined by the identity of the first
collision.  When the piston is initially located at $x_2=1/2$ and the two
particles approach with equal and opposite velocities, the piston is driven
to the right wall if the first collision occurs with its right neighbor and
{\it vice versa}.  The light particle that first hits the piston then
collides earliest with the wall and begins cooling earlier.  This fact leads
to the piston eventually compressing the particle that experiences the first
collision with a wall.

\subsection{The log-periodic state}
\label{logp}

Numerically, we find that the three-particle system asymptotically falls into a
log-periodic state--where the piston undergoes small-amplitude oscillations
with a constant period in $\ln t$---for almost all initial conditions.  In
this state, one of the light particles is trapped in a small gap between the
piston and the wall (Fig.~\ref{logfinal}), while the other light particle has
most of the energy and travels over almost the entire interval.

During these oscillations, the light particle that is compressed by the
piston performs a sequence of violent rattlings each time the piston
approaches the wall that eventually reflect the piston from the wall (top
panel in Fig.~\ref{logfinal}).  The piston then collides with the other light
particle whose energy is nearly equal to that of the entire system and whose
momentum is comparable in magnitude to that of the piston.  After this
collision, the piston is reflected back toward the nearer wall and the
rattling sequence with the trapped light particle begins anew.

Generally this long-time state has a one-cycle periodicity in which the
position of the piston recurs at each maximum of its oscillation cycle
(Fig.~\ref{logfinal}).  However, for piston mass $m_2$ less that a
$r$-dependent threshold mass $\mu_t(r)$, we empirically find that the
asymptotic state can be a two-cycle, three-cycle, etc., with lower cycles
more likely to occur than high cycles.  Conversely, for $m_2$ greater than an
upper threshold
\begin{equation}
  \mu_c(r)=\frac{(1+r)(1+r+4\sqrt{r})+4r}{(1-r)^2}~,
\label{mcritbis}
\end{equation}
inelastic collapse occurs, where the piston ultimately sticks to a wall (see
the Appendix for the derivation of $\mu_c$).  For the purposes of the present
discussion, we are interested in the case where $m_2$ is in the range
$[\mu_t(r),\mu_c(r)]$ so that the system falls into a one-cycle log-periodic
state.

This state may be characterized by the relaxation time $\tau(m_2,r)$ until
the piston settles into the log-periodic motion and the amplitude,
$A(m_2,r)$, and period on a logarithmic time scale, $\Delta(m_2,r)$, of the
ensuing oscillations.  The latter is defined via
$t_{k+1}=\textrm{e}^{\Delta}t_k$, where $t_k$ and $t_{k+1}$ are the times for
two consecutive maxima of $x_2(t)$ in the final state (bottom panel in
Fig.~\ref{logfinal}).

\begin{figure}[t] 
 \vspace*{0.cm}
 \includegraphics*[width=0.45\textwidth]{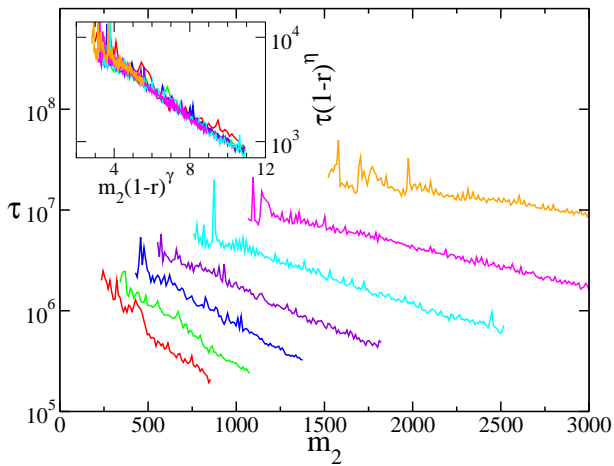}
\caption{(Color online) Plot of $\tau$ as a function of $m_2$ on a
  semilogarithmic scale, for restitution coefficients $r=0.875$, 
  0.8875, 0.9, 0.9125, 0.925, 0.9375, and 0.95 (bottom to top).   The inset 
  shows the data collapse of $\tau$ for different values of $r$ using
  $\gamma=2.1$ and $\eta=2.6$. 
  \label{relaxt}}
\end{figure}

\begin{figure}[t] 
  \vspace*{0.cm} \includegraphics*[width=0.45\textwidth]{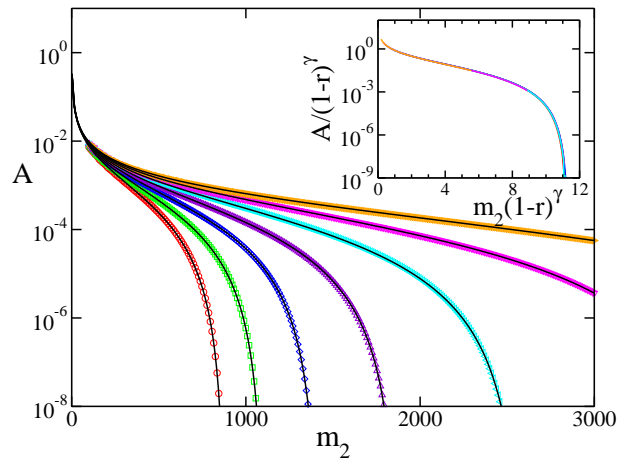}
  \caption{(Color online) The amplitude of the log-periodic oscillations,
    $A$, as a function of the piston mass on a semilogarithmic scale for the
    same $r$ values as in Fig.~\ref{relaxt} (data shifting to the right for
    increasing $r$).  The curves are the predictions from Eq.~(\ref{xpeq}),
    based on an effective restitution coefficient picture (see text).  Inset:
    Data collapse of $A$ for different values of $r$ and $\gamma=2.1$.
    \label{ampli}}
\end{figure}

Figure \ref{relaxt} shows the relaxation time $\tau$ as a function of $m_2$ for
$m_2$ in the range $[\mu_t,\mu_c]$ for representative values of $r$.  We
expect that the oscillatory regime is reached more quickly for larger $m_2$
since energy is more quickly dissipated when the piston is heavier, as
confirmed by the data.  We also find that $\tau$ decays exponentially with
$m_2$, that is, $\tau\sim \exp[-m_2/\mu(r)]$, with a characteristic mass
scale $\mu(r)$ that is nearly equal to the threshold mass $\mu_t(r)$.  Both
$\mu$ and $\mu_t$ numerically scale as $(1-r)^{-\gamma}$ for $r$ close to 1,
with $\gamma\approx 2.1$.  This is slightly larger than the anticipated
exponent value of 2 that is based on the hypothesis that there should be only
one characteristic mass that scales as $\mu_c\sim (1-r)^{-2}$ in the limit
$r\to 1$ from Eq.~(\ref{mcritbis}).  We attribute the discrepancy in $\gamma$
to corrections to scaling; the largest restitution coefficient $r=0.95$ that
is practical to study is still not very close to 1.

When the piston is in the log-periodic state, the amplitude $A$ is a
monotonically decreasing function of $m_2$ and vanishes as $m_2 \to
\mu_c(r)$, signaling the onset of inelastic collapse (Fig.~\ref{ampli}).  The
logarithmic period of the oscillations $\Delta$ (not shown) scales
approximately as $\Delta(m_2,r)\approx(1-r)$ and depends weakly on $m_2$.

Because these three characteristics of the oscillations---$\tau$, $A$, and
$\Delta$---seem to be governed by the same mass scale, we anticipate that
data collapse will occur.  Empirically, we find that these quantities are
consistent with the scaling forms
\begin{eqnarray}
\tau(m_2,r) &\sim&(1-r)^{-\eta}~\Phi_{\tau}\big[m_2(1-r)^{\gamma} \big]\,, \nonumber \\
A(m_2,r) &\sim&(1-r)^{\gamma}~\Phi_{A}\big[m_2(1-r)^{\gamma} \big]\,,\label{scaling} \\
\Delta(m_2,r) &\sim & (1-r)~\Phi_{\Delta}\big[m_2(1-r)^{\gamma} \big]\,, \nonumber
\end{eqnarray}
with $\gamma\approx 2.1$.  Additionally, $\eta\approx 2.6$ is an apparently
independent exponent that characterizes the relaxation time $\tau$, while
$\Phi_\tau$, $\Phi_A$, and $\Phi_\Delta$ are scaling functions. The insets to 
Figs.~\ref{relaxt} and \ref{ampli} show that the data collapse for $\tau$ and 
$A$ is quite good. 

\begin{figure}[ht] 
 \vspace*{0.cm}
\includegraphics*[width=0.45\textwidth]{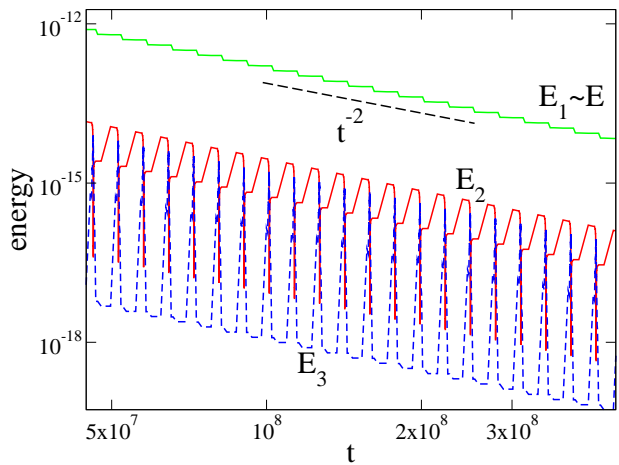}
\caption{(Color online) Energy of each particle as a function of time on a
  double logarithmic scale for the system depicted in Fig.~\ref{logosc}.
  \label{totenergy}}
\end{figure}

Although the asymptotic state of the system is periodic on a logarithmic time
scale, we emphasize that the total energy of the system, $E(t)$, continues to
dissipate due to inelastic collisions with the walls.  At a coarse-grained
level, we recover Haff's law \cite{H} $E(t)\sim t^{-2}$ (upper curve in
Fig.~\ref{totenergy}), as expected.  Notice that for the specific example
being studied (in which the piston compresses particle 3), $E(t)\approx
E_1(t)$.  At a finer time scale, however, $E(t)$ undergoes a sequence of
steps and almost constant plateaus.  The largest drop in energy occurs when
the more energetic light particle collides with the wall, while the rattling
dynamics between the piston and the other light particle leads to a small
decrease in the energy of the system.

\section{MACROSCOPIC DESCRIPTION}

We can understand the initial instability of the piston in terms of
macroscopic equations of motion \cite{H,BRV05}.  The macroscopic approach
given here ostensibly applies for any value of the restitution coefficient
$r$ for fixed piston mass $m_2$, or equivalently for any $m_2$ for fixed $r$.
In particular, this approach correctly describes the initial instability of
the inelastic piston for any value of the parameters $r$ and $m_2$.  The only
feature that the macroscopic approach fails to describe is the final state
for piston masses below the critical mass, $m_2<\mu_c$.

According to the macroscopic description, the energies of the light particles
change in time by two processes. First, the energy decreases due to inelastic
collisions with the walls.  This cooling is macroscopically described by
Haff's law, in which the energy change is proportional to the particle energy
and the number of collisions per unit time.  Thus ${ d}E_i(t) \vert_{\rm
  coll} = -E_i(t) (1-r^2) n_{\rm coll}(t){ d}t$, with $i=1,3$.  The
collision rate may be approximated by $n_{\rm coll}(t)\approx
\sqrt{2E_i(t)}/\ell_i(t)$, where $\sqrt{2E_i(t)}$ is the thermal velocity and
$\ell_i(t)$ is the length of the region available for particle $i$; thus
$\ell_1(t)=x_2(t)$ and $\ell_3(t)=1-x_2(t)$.  On the other hand, the energies
of the light particles also change because of compression or expansion by the
piston.  The macroscopic equation describing this process is ${ d}E_i(t)
\vert_{\rm piston} = -P_i{ d}\ell_i$, $i=1,3$, where $P_i$ is the pressure
exerted on particle $i$, and ${ d}\ell_i$ is the length change

Assuming the ideal gas law $P_i=T_i/\ell_i$ (Boltzmann's constant is set to
1), and writing $T_i=2E_i$, we obtain
\begin{eqnarray}
\frac{{ d}E_1}{{ d}t} & = & - 2E_1\frac{v_2}{x_2} -
\sqrt{2}(1-r^2)\frac{E_1^{3/2}}{x_2}~, \label{macro1} \\
\frac{{ d}E_3}{{ d}t} & = & 2E_3\frac{v_2}{1-x_2} -
\sqrt{2}(1-r^2)\frac{E_3^{3/2}}{1-x_2}~. \label{macro2}
\end{eqnarray}
This is essentially the approach of Haff \cite{H} for the inelastic gas, and
it was also adopted for the inelastic many-particle piston problem of Brito
\emph{et al.}\ \cite{BRV05}. The force exerted on the piston is given by the
pressure difference, $P_1-P_2$, so that the macroscopic equation of motion
for the piston is
\begin{equation}
m_2\frac{{ d}^2x_2}{{ d}t^2} = \frac{2E_1}{x_2} - \frac{2E_3}{1-x_2}~.
\label{macro3}
\end{equation}
Equations (\ref{macro1})--(\ref{macro3}) describe the evolution of the three-particle
system on a coarse-grained time scale and they are the analogs of the
equations derived in Ref.~\cite{BRV05} for the many-body inelastic piston
problem.  

A particular solution to these macroscopic equations is symmetric cooling of
both light particles, $E_1(t)=E_3(t)= E_0\big[1+\sqrt{2E_0}(1-r^2)~t
\big]^{-2}$, with $E_0$ the initial light particle energy, while the piston
remains at $x_2(t)=1/2$.  However, linear perturbation analysis shows that
any small disturbance from symmetry grows and the piston is driven toward one
of the walls, with an oscillatory modulation that is periodic on logarithmic
time scale \cite{BRV05}.

A typical piston trajectory that is obtained by numerically solving
Eqs.~(\ref{macro1})--(\ref{macro3}) with a slightly asymmetrical state is
shown in Fig.~\ref{logosc}.  The numerical solution to the macroscopic
equations and the simulation results for the three-particle system are extremely
close over the time range $10^3 < t < 10^5$.  However, after approximately
$10^5$ time steps (for the case $m_2=100$ and $r=0.9$), the macroscopic
equations predict that inelastic collapse occurs, after which the piston
sticks to one of the walls \cite{BRV05}.  In contrast, for the three-particle
system, the piston localizes near one wall but continues to undergo
small-amplitude, nearly regular oscillations on a logarithmic time scale
(Fig.~\ref{logfinal}).

To help understand this discrepancy between the macroscopic approach and the
simulations results for the three-particle system in the long-time limit, it is
helpful to reconsider the elastic case $r=1$.  Here Eqs.~(\ref{macro1}) and
(\ref{macro2}) can be immediately integrated, and substituting the results of
these integrations into (\ref{macro3}) gives
\begin{equation}
m_2\frac{{d}^2x_2}{{d}t^2} = \frac{A_1}{x_2^3} - 
\frac{A_3}{(1-x_2)^3}~, \nonumber
\end{equation}
where $A_{1,3}$ are constants.  This equation of motion describes the
oscillations of a particle in the effective potential well $V_{\rm eff}(x)=
\frac{1}{2}[A_1x^{-2} + A_3(1-x)^{-2}]$.  This effective potential can be
derived rigorously in the limit $m_2\to\infty$ (see \cite{Sinai} and also the
Appendix of HR).  Thus the long-time extreme excursions in the elastic
system, which are not described by the effective potential, appear to
stem from the finiteness of the piston mass.

By analogy, we anticipate that the macroscopic equations
(\ref{macro1})--(\ref{macro3}) should describe the final state for the
inelastic piston in the $m_2\to\infty$ limit.  On the other hand, the
log-periodic state emerges only when the piston mass is finite.  This feature
seems to play a parallel role as in the elastic system, in that departures
from the predictions of the macroscopic equation arise only when the piston
mass is finite.

\section{EFFECTIVE RESTITUTION COEFFICIENT}

To understand the properties of the log-periodic oscillations, we map the
three-particle system onto an equivalent two-particle system, from which the basic
characteristics of the log-periodic state follow.  The first step is to
determine the net effect of the sequence of rattling collisions between the
piston and a light particle as the piston approaches a wall and is ultimately
reflected.  We show in the Appendix that this collision sequence is
equivalent to a 1-body problem in which the piston is reflected from the wall
with an effective restitution coefficient $r_{\rm eff}(m_2,r)$ that is
smaller than the bare restitution coefficient $r$.

\begin{figure}[b] 
 \vspace*{0.cm}
\includegraphics*[width=0.45\textwidth]{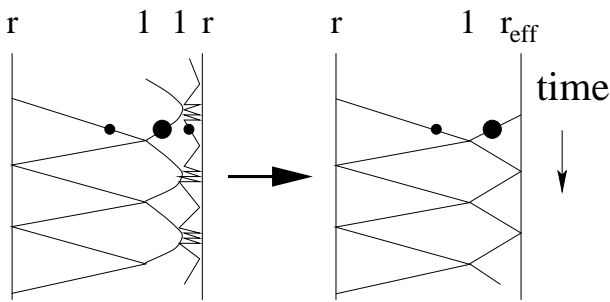}
\caption{Schematic space-time diagram of the particle trajectories in the
  log-periodic state (left) and the effective trajectories (right).
  \label{reff-pict}}
\end{figure}

Next, we exploit the symmetry breaking, in which the piston localizes near
one wall, to reduce the initial three-body problem into an effective two-body
problem that consists of the piston and one light particle.  In this reduced
system, the piston collides inelastically with the wall with restitution
coefficient $r_{\rm eff}$, while the light particle collides elastically with
the piston and inelastically with the other wall with restitution coefficient
$r$ (Fig.~\ref{reff-pict}).  Using this equivalence, we will determine the
properties of the log-periodic state.

\begin{figure}[t] 
\vspace*{0.cm}
\includegraphics*[width=0.45\textwidth]{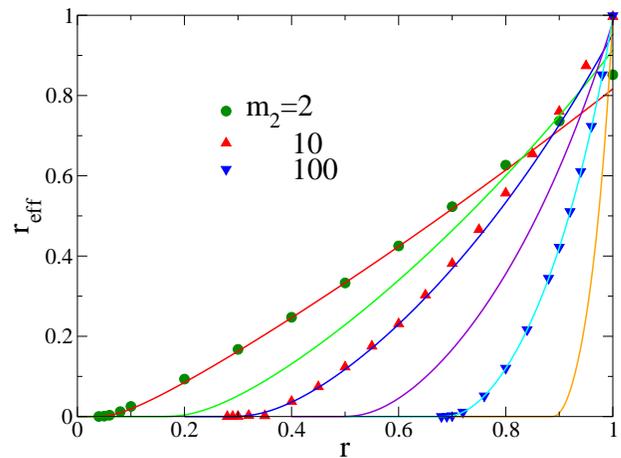}
\caption{(Color online) The effective restitution coefficient as a function of 
  $r$ for piston masses $m_2=2$, 5, 10, 30, 100, and 1000.  The initial
  velocities are $(v_1,v_2)=(0,-1)$.  The curves are the theoretical predictions from
  Eq.~(\ref{reffeq}), and the symbols correspond to simulation results.
  \label{reff}}
\end{figure}

For the initial step of determining the effective restitution coefficient as
a function of $r$ and $m_2$, the calculational details are given in the
Appendix and the final result for $r_{\rm eff}(m_2,r)$ is
quoted in Eq.~(\ref{reffeq}).  As shown in Fig.~\ref{reff}, $r_{\rm eff}$
decreases as $r$ decreases and goes to zero as $r$ approaches a critical
value $r_c(m_2)$, quoted in Eq.~(\ref{rcrit}), that signals inelastic
collapse.  When $r<r_c$, the effective restitution coefficient is zero, and
the result of the rattlings between the piston and the intervening light
particle is inelastic collapse.  For fixed $r$, notice also that $r_{\rm
  eff}$ decreases rapidly as $m_2$ is increased.

With this effective restitution coefficient equivalence, we now reduce the
original three-particle system to the equivalent two-particle system.  Without loss
of generality, we assume that the piston is close to the wall at $x=1$.  The
effective system then consists of a light particle at $x_1$ and the piston at
$x_2$, with $0<x_1<x_2<1$.  For sufficiently large piston mass, the sequence
of collisions consists of: (i) the piston making an effective collision with
the wall with restitution coefficient $r_{\rm eff}$, (ii) the second light
particle undergoing a bare inelastic collision with the other wall, and (iii)
an elastic particle-piston collision, with each of these steps being
non-overlapping.  From the collision rules for each of these steps [see
Eqs.~(\ref{12})--(\ref{W})], the new velocities after each such collision
sequence are given in terms of the incoming velocities by
\begin{eqnarray}
\label{vL}
\left(\! \begin{array}{c}{\displaystyle v_1'}\\ {\displaystyle
      v_2'}\end{array}\! \right) \!=\!
\left( \! \begin{array}{cc}
{\displaystyle \frac{m_2-1}{M}\,\,r} &{\displaystyle -\frac{2m_2}{M}\,\,r_{\rm eff}} \\ \\
{\displaystyle -\frac{2}{M}\,\,r} &{\displaystyle -\frac{m_2-1}{M}\,\,r_{\rm eff}} 
\end{array} \! \right)\!\!
\left( \!\!\begin{array}{c}
{\displaystyle v_1} \\ {\displaystyle v_2} \end{array} \!\!\right)  
\equiv
{\bf L\,\,v}\,,
\end{eqnarray}
where $M=1+m_2$, and $v_1,v_1'<0$ and $v_2,v_2'>0$.  The velocity vector
after $n$ such cycles is ${\bf v}^{(n)}={\bf L}^n {\bf v}$.  Diagonalizing
${\bf L}$, we find (see also the Appendix)
\begin{eqnarray}
v_1^{(n)} & = & \frac{\Lambda_+^n(\Lambda_--a) - \Lambda_-^n(\Lambda_+-a)}
{\Lambda_--\Lambda_+}~v_1 \nonumber \\ 
& + & \frac{b(\Lambda_-^n-\Lambda_+^n)}{\Lambda_--\Lambda_+}~v_2~, \label{v1n}\\
v_2^{(n)} & = & \frac{(\Lambda_+^n-\Lambda_-^n)(\Lambda_+-a)(\Lambda_--a)}
{b(\Lambda_--\Lambda_+)}~v_1 \nonumber \\
& + & \frac{\Lambda_-^n(\Lambda_--a) - \Lambda_+^n(\Lambda_+-a)}
{\Lambda_--\Lambda_+} v_2~, \label{v2n}
\end{eqnarray}
where $a=r(m_2-1)/M$ and $b=-2r_{\rm eff}m_2/M$ are the elements in the first
row of ${\bf L}$, and $\Lambda_{\pm}$ are the eigenvalues of matrix ${\bf
  L}$,
\begin{equation}
\Lambda_{\pm} = \frac{m_2-1}{2M}(r-r_{\rm eff})\left( 1 \pm \sqrt{1+
\frac{4rr_{\rm eff}M^2}{(m_2-1)^2(r-r_{\rm eff})^2} } \right).
\label{eigenL}
\end{equation}
Both eigenvalues are real, with $\Lambda_-<0$, $\Lambda_+>0$,  and 
$\vert\Lambda_-\vert<\vert\Lambda_+\vert <1$. 

We test this effective description of the collision dynamics by comparing the
exact piston trajectory in the three-particle system for a given $m_2$ and $r$
with the piston trajectory in the reduced two-particle system.  After shifting
the effective trajectory by an overall phase factor, both systems have
visually indistinguishable periodic behavior, thus confirming the validity of
the coarse-grained approach.

We now determine the relation that $v_1^{(n)}$ and $v_2^{(n)}$ must satisfy
for the effective two-particle system to be in a log-periodic state.  For
such a periodicity, successive collisions between particle 1 and the piston
must occur at the same position $x_0$ for all $n$.  Thus the time $\delta
t_i^{(n)}$ for particle $i$ to go from $x_0$ to its respective wall and
return to $x_0$ in the $n^{\rm th}$ cycle must be the same for both
particles.  That is,
\begin{eqnarray*}
\delta t_1^{(n)}=\frac{x_0}{\vert v_1^{(n)}\vert}\frac{1+r}{r}=
\delta t_2^{(n)}=\frac{1-x_0}{\vert v_2^{(n)}\vert}\frac{1+r_{\rm eff}}{r_{\rm eff}}
\equiv \delta t^{(n)}\,.
\end{eqnarray*}
Thus
\begin{equation}
\frac{\vert v_2^{(n)}\vert}{\vert v_1^{(n)}\vert} = 
\frac{1-x_0}{x_0}~\frac{r(1+r_{\rm eff})}{r_{\rm eff}(1+r)}
\label{condi}
\end{equation}
is a constant that is independent of $n$ in the log-periodic state.
Therefore $\vert v_2^{(n+1)}\vert/\vert v_1^{(n+1)}\vert= \vert
v_2^{(n)}\vert/\vert v_1^{(n)}\vert$.  Then using ${\bf v}^{(n+1)}={\bf L
  v}^{(n)}$, we express ${\bf v}^{(n+1)}$ in terms of ${\bf v}^{(n)}$ and
thereby obtain
\begin{equation}
\label{velratio}
\frac{\vert v_2^{(n)}\vert}{\big\vert v_1^{(n)}\vert} = \frac{(1+m_2)\Lambda_+ - (m_2-1)r}{2m_2r_{\rm eff}}~,
\end{equation}
where $\Lambda_+(m_2,r)$ is the larger eigenvalue of $\mathbf{L}$.  Comparing
Eqs.~(\ref{condi}) and (\ref{velratio}) finally yields
\begin{equation}
x_0=\left(1+\frac{(1+r)\big[(1+m_2)\Lambda_+- (m_2-1)r\big]}
{2m_2r(1+r_{\rm eff})}\right)^{-1}~,
\label{xpeq}
\end{equation}
and the amplitude of the piston oscillations is then $A(m_2,r)=1-x_0$.  This
result agrees with the simulation results shown in Fig.~\ref{ampli}, even
close to inelastic collapse.

We may also compute the logarithmic period of the piston oscillations.  Since
$\delta t^{(n+1)}/\delta t^{(n)} = \vert v_2^{(n)}\vert/\vert
v_2^{(n+1)}\vert$, we express $\vert v_2^{(n+1)}\vert$ in terms of $\vert
v_1^{(n)}\vert$ and $\vert v_2^{(n)}\vert$ from Eq.~(\ref{v2n}), and then
use Eq.~(\ref{velratio}) to obtain
\begin{equation}
\frac{\delta t^{(n+1)}}{\delta t^{(n)}} = \frac{\Lambda_+ - \dfrac{m_2-1}{1+m_2}r}
{r_{\rm eff} \Big(r-\dfrac{m_2-1}{1+m_2}\Lambda_+ \Big)} >1~.
\label{deltaratio}
\end{equation}
Thus in the log-periodic state $\delta t^{(n)}$ grows exponentially in the
number of cycles $n$, as seen in our simulations.  From the logarithmic
period $\Delta(m_2,r)$ introduced in Sec.~\ref{logp}, we have $t_{n-1}=
e^{-\Delta} t_n$, so that $\delta t^{(n)}= t_n(1-e^{-\Delta})$.  This
relation then gives $\Delta=\ln(t_{n+1}/t_n)=\ln(\delta t^{(n+1)}/\delta
t^{(n)})$.  The agreement between this prediction for $\Delta$ with
simulation results (not shown) is again extremely good.

Finally, the robustness of the log-periodic state can be understood in simple
terms.  Starting with an arbitrary (not log-periodic) initial state, it is
easy to show from Eqs.~(\ref{v1n}) and (\ref{v2n}) that both $v_1^{(n)}$ and
$v_2^{(n)}$ converge exponentially quickly with $n$ to a state where the
ratio $\vert v_2^{(n)}\vert/\vert v_1^{(n)}\vert$ satisfies the condition
(\ref{velratio}) that signals log-periodicity.  This convergence occurs
because for $r_{\rm eff}<r$, $(\Lambda_-/\Lambda_+)^n$ quickly goes to zero
as $n$ increases.  In this sense, the log-periodic state is an attractor of
the dynamics.

\section{SUMMARY}

We investigated the dynamics of a three-particle system on the unit interval in
which a massive particle (corresponding to a piston) lies between 2 light
particles.  The particles collide elastically with the piston, but
inelastically with the walls.  This toy model is meant to mimic the behavior
of the inelastic piston problem in which a massive piston separates two
inelastic gases, each of which contains many particles.  The dynamics of this
many-body problem is extremely rich.  The piston moves nonmonotonically at
early times and correspondingly the response of the two gases is also
nonmonotonic.  Eventually there is an inelastic collapse in which one of the
gases is compressed into a solid by the piston.

One of the motivations for our study of the three-particle system was to capture
some of the intriguing phenomenology of the many-particle inelastic piston
problem.  A new feature of the three-particle system, however, is that the piston
settles into a log-periodic state at long times over a wide range of
restitution coefficients, in which the period is constant on a logarithmic
time scale.  The characteristics of this log-periodic state can be understood
in terms of a simple effective picture in which the rattling collision
sequence between the piston and the trapped light particle is replaced by an
effective inelastic collision between the piston and the wall, with effective
restitution coefficient $r_{\rm eff}<r$.  This equivalence provides a
satisfyingly complete account of the log-periodic state.  Finally, it should
be noted that the log-periodic behavior is a consequence of the finiteness of
the piston mass.  As $m_2$ increases, the amplitude of the oscillations
decreases and as $m_2\to\infty$ the inelastic collapse of the many-particle
inelastic piston problem is recovered.

\acknowledgments{We thank R. Brito for collaboration during the initial
  stages of this project.  S.R. thanks NSF grant No. DMR0227670 (BU) and DOE grant
  No. W-7405-ENG-36 (LANL) for financial support.  P.I.H. acknowledges support from
  Spanish MEC, and thanks LANL and CNLS for hospitality of during part of
  this project.}

\appendix*

\section{CALCULATION OF THE EFFECTIVE RESTITUTION COEFFICIENT}
\label{app-reff}

We use a matrix approach to compute the effective restitution coefficient
$r_{\rm eff}$ that describes the final velocity of the piston at the end of
the rattling collisions as a function of $m_2$ and the bare restitution
coefficient $r$.  Without loss of generality, we assume that the piston
compresses particle 1 which then undergoes the rattling collision sequence.
For concreteness, the light particle is taken to be at rest at $x_1>0$ with a
wall at $x=0$.  A massive particle (the piston) approaches the light particle
from the right with $v_2=-1$.  Collisions between particle 1 and the wall at
$x=0$ are inelastic, with restitution coefficient $r$, while 1-2 collisions
are elastic.  After the rattling collision sequence ends, the piston recedes
from the wall with velocity $v_2'=-v_2\,r_{\rm eff}$.

The velocities after each collision are given in terms of the velocities
before the collision by
\begin{eqnarray}
v_1' & = & \frac{1-m_2}{M}\,\, v_1+ \frac{2m_2}{M}\,\, v_2 \, , \nonumber \\
v_2' & = & \frac{2}{M} \,\,v_1 +\frac{m_2-1}{M}\,\, v_2 \, ,
\label{12}
\end{eqnarray}
for the 1--2 collisions, and
\begin{eqnarray}
 v_{1}' & = & -v_{1}\,r \, , \nonumber \\ 
 v_{2}' & = & v_{2} \, ,
\label{W}
\end{eqnarray}
for the wall collision, where $M=1+m_2$.  Thus the combined effect of a 1-2 and an ensuing
particle-wall collision is given by the composition of the two
transformations implicit in Eqs.~(\ref{12}) and (\ref{W}).  Therefore
\begin{eqnarray}
\label{matrix}
\left(\! \begin{array}{c}{\displaystyle v_1'}\\ {\displaystyle
      v_2'}\end{array}\! \right)\!\!=\!\!
\left( \begin{array}{cc}
{\displaystyle -\frac{1-m_2}{M}\,\,r} &{\displaystyle -\frac{2m_2}{M}\,\,r} \\ \\
{\displaystyle\frac{2}{M}} &{\displaystyle \frac{m_2-1}{M}} \end{array} \right)
\!\!\left(\!\! \begin{array}{c}
{\displaystyle v_1} \\ {\displaystyle v_2} \end{array} \!\! \right)
\equiv
{\bf M\,v} \, .
\end{eqnarray}

The velocity vector after $n$ such cycles is given by 
\begin{equation}
{\bf v}^{(n)}={\bf  M}^n {\bf v}^{(0)}\,, \quad {\rm with} \quad {\bf v}^{(0)}= \left({~0\atop -1}\right)\, . \nonumber
\end{equation}
The collision sequence ends when the velocities of the two particles after $n$
cycles satisfy $v_1^{(n)}-v_2^{(n)}<0$,
corresponding to the two particles receding from the wall with the piston
moving faster than the light particle.  We define this situation as
``escape'' of the piston.  The number of collisions for escape to occur is
given by the smallest value of $n$ that leads to the above conditions on the
outgoing velocities.  The effective restitution coefficient is then given by
$v_2^{(n)}$ when $n$ equals its value at escape.
 
To determine the threshold value of $n$, we use the fact that (see, 
  e.g., \cite{Arf})
\begin{eqnarray}
\label{diag}
\bf {M^n\,\, v}= {\bf S \,\, M_{\rm diag}^n \,\, S^{-1}\, v},
\end{eqnarray}
where ${\bf S}$ is the similarity matrix that diagonalizes {\bf M}, and ${\bf
  M_{\rm diag}= S^{-1} \,\, M^n \,\, S}$ is the diagonalized form of the
transformation matrix.  The eigenvalues of {\bf M} are $\lambda_\pm =
(T\pm\sqrt{T^2-4D})/2$, where $T=(m_2-1)(1+r)/M$ is the trace and $D=r$ is
the determinant of {\bf M},
\begin{equation}
\lambda_{\pm} = \frac{(m_2-1)(1+r)}{2M}\left(1 \pm 
\sqrt{1-\frac{4rM^2}{(m_2-1)^2(1+r)^2}} \right)~.
\label{eigenval}
\end{equation}
Consequently the similarity transformation matrix is 
\begin{eqnarray*}
{\bf S} =\left( \begin{array}{cc}
1 &1 \\ \\{\displaystyle \frac{\lambda_+-a}{b}} 
& {\displaystyle\frac{\lambda_--a}{b}} \end{array} \right),
\end{eqnarray*}
where $a=r(m_2-1)/M$ and $b=-2m_2r/M$ are the elements of the first row of
{\bf M}, i.e., the matrix {\bf S} consists of the eigenvectors of {\bf
  M} arranged columnwise.  Consequently ${\bf S}^{-1}= {\rm \vert
  S\vert}^{-1}{\bf S}^\dag$, where ${\rm \vert S\vert}$ is the determinant of
{\bf S}, and ${\bf S}^\dag$ is its transpose.

Assembling these results, the velocity after $n$ cycles (and $2n$ individual
collisions) is
\begin{eqnarray}
\label{v-matrix}
{\bf v}^{(n)} \!=\! \left( \!\! \begin{array}{c}
{\displaystyle b\frac{\lambda_+^n-\lambda_-^n}{\lambda_--\lambda_+}} \\ \\ {\displaystyle 
\frac{\lambda_+^n(\lambda_+-a)}{\lambda_--\lambda_+} - 
\frac{\lambda_-^n(\lambda_--a)}{\lambda_--\lambda_+}} \end{array}\!\! \right)
\!\equiv\! \left( \!\! \begin{array}{c}  v_1^{(n)} \\ \\ v_2^{(n)}
  \end{array}\!\! \right)\!.
\label{veloc}
\end{eqnarray}
In the case where escape of the piston requires $n+1$ particle-particle
collisions and $n$ particle-wall collisions, we should multiply the
transformation matrix ${\bf M}^n$ on the left by the matrix defined by
Eq.~(\ref{12}) to account for this last particle-particle collision.
However, to compute only the final velocity of particle 2, it suffices to
calculate ${\bf v}^{(n+1)}$ from Eq. (\ref{veloc}).

Depending on the sign of the discriminant $T^2-4D$, the eigenvalues
$\lambda_{\pm}$ can be real or complex.  For $r$ greater than a threshold
value $r_c(m_2)$, $T^2<4D$.  Thus $\lambda_{\pm}$ are complex conjugates (note, 
however, that ${\bf v}^{(n)}$ has always real components).  At the threshold, $T^2=4D$, 
leading to $\lambda_+=\lambda_-$, so that ${\bf v}^{(n)}$ is undetermined.  This
indeterminacy signals inelastic collapse: for $r<r_c(m_2)$ there is an
infinite number of collisions in a finite time, and $v_1^{(n)}-v_2^{(n)}>0$
$\forall \, n$.  The condition $T^2=4D$ gives the critical restitution
coefficient for inelastic collapse:
\begin{equation}
r_c(m_2) = \frac{(1+m_2)(1+m_2-4\sqrt{m_2})+4m_2}{(m_2-1)^2}.
\label{rcrit}
\end{equation}
Notice that $r_c\sim 1-4/\sqrt{m_2}$ in the limit of large $m_2$.
Equivalently, the condition $T^2=4D$ defines a critical mass $\mu_c(r)$, such
that inelastic collapse occurs for $m_2>\mu_c(r)$.  We now find
\begin{equation}
\mu_c(r)=\frac{(1+r)(1+r+4\sqrt{r})+4r}{(1-r)^2}.
\label{mcrit}
\end{equation}
Note that for $r$ close to 1, $\mu_c(r) \sim 16(1-r)^{-2}$.

For $r>r_c(m_2)$ the piston eventually escapes with velocity $v_2^{(n_0)}$,
where $n_0$ is the number of cycles until escape.  To determine $n_0$, define
$f(n)\equiv v_1^{(n)}-v_2^{(n)}$.  Initially $f(0)=1$, and $f(n)$ decreases 
as $n$ increases and eventually changes sign. Next, we define the real variable $z$ by 
the condition $f(z)=0$.  From Eq.~(\ref{v-matrix}),
\begin{equation}
\label{fz}
f(z)=\frac{b(\lambda_+^z- \lambda_-^z) -
\lambda_+^z(\lambda_+-a) + \lambda_-^z(\lambda_--a)}{\lambda_--\lambda_+} =0\,.
\end{equation}
Since $\lambda_{\pm}$ are complex conjugates, we write $\lambda_{\pm}=
Q\,e^{\pm i\beta}$ so that Eq.~(\ref{fz}) becomes, using
$\lambda_+^z-\lambda_-^z=2iQ^z\sin(z\beta)$ and $a+b=-r$,
\begin{equation}
\label{f-soln}
r\sin(z\beta)+Q\sin[(z+1)\beta]=0\,,
\end{equation}
with solutions
\begin{equation}
\label{nmax}
z(k)=\frac{1}{\beta}\Big[k\pi-\tan^{-1}\left(\frac{Q\sin\beta}{r+Q\cos\beta}\right)\Big]\,,
\end{equation}
where $k$ can be any integer number. The first solution that has a physical
meaning (i.e., $z>1$) corresponds to $k=1$, so $z=z(1)$.  The number of
collision cycles before escape is thus $n_0=\lceil z\rceil$, where $\lceil
z\rceil$ is the next integer larger than $z$.  The escape velocity is
$v_2^{(n_0)}$ and $r_{\rm eff}=v_2^{(n_0)}$.  However, for large enough
piston mass, the number of collisions before escape is typically large, and
we can approximate $n_0$ by $z$.  Hence, we finally obtain for the effective
restitution coefficient,
\begin{equation}
r_{\rm eff}(m_2,r) = \frac{\lambda_+^{z}(\lambda_+-a) - 
\lambda_-^{z}(\lambda_--a)}{\lambda_--\lambda_+}~,
\label{reffeq}
\end{equation}
with $z$ given by Eq.~(\ref{nmax}) with $k=1$.  A plot of $r_{\rm eff}$ as a
function of $r$ is given in Fig.~\ref{reff}.


\begin{thebibliography}{99}
  
\bibitem{HR1} P. I. Hurtado and S. Redner, Phys. Rev. E {\bf 73}, 016136 (2006).
  
\bibitem{C60} H. B. Callen, {\it Thermodynamics} (J. S. Wiley \& Sons, New
  York, 1960).

\bibitem{L99} E. H. Lieb, Physica A {\bf 263}, 491 (1999).

\bibitem{KVM00} E. Kestemont, C. Van den Broeck, and M. Malek Mansour,
  Europhys.\ Lett.\ {\bf 49}, 143 (2000).

\bibitem{GPL02} C. Gruber, S. Pache, and A. Lesne, J. Stat.\ Phys.\ {\bf
    108}, 669 (2002).

\bibitem{CLS02} N. I. Chernov, J. L. Lebowitz, and Ya. G. Sinai, Russ.\ Math.\ Surveys\
 {\bf 57}, 1045 (2002). 

\bibitem{GZ} G. Galperin and A. Zemlyakov, {\it Mathematical Billiards} (Nauka, Moscow, 1990) 
(in Russian).
  
\bibitem{KT} V. V. Kozlov and D. V. Treshsh\"ev, {\it Billiards: A Generic
    Introduction to the Dynamics of Systems with Impacts} (Amererican
    Mathematical Society, Providence, RI, 1991).
    
\bibitem{T} S. Tabachnikov, {\it Billiards} (Soci\'et\'e Math\'ematique de
    France, Amererican Mathematical Society, Providence, RI, 1995).
  
\bibitem{G} E. Gutkin, J.  Stat.\ Phys.\ {\bf 81}, 7--26 (1996).

\bibitem{R04} S. Redner, Am.\ J. Phys.\ {\bf 72}, 1492 (2004).  

\bibitem{BRV05} R. Brito, M. J. Renne, and C. Van den Broeck, Europhys.\
  Lett.\ {\bf 70}, 29 (2005).
  
\bibitem{general} See, e.g., S. McNamara and W. R. Young, Phys.\ Fluids
  A {\bf 4}, 496 (1992); Phys.\ Rev.\ E {\bf 53},
  5089 (1996); I. Goldhirsch and G. Zanetti, Phys.\ Rev.\ Lett.\ {\bf 70},
    1619 (1993); P. Deltour and J.-L. Barrat, J. Phys. (Paris) {\bf 7},
  131 (1997).
  
\bibitem{H} P. Haff, J. Fluid Mech.\ {\bf 134}, 401 (1983).

\bibitem{Sinai} Ya. G. Sinai, Theor. Math. Phys. {\bf 121}, 1351 (1999).

\bibitem{Arf} See, e.g., G. Arfken and H. J. Weber, {\it Mathematical
    Methods for Physicists}, 5th ed. (Academic Press, New York, 2000).


\end{thebibliography}
\end{document}